# Design and Qualification of an Airborne, Cosmic Ray Flux Measurement System

Sanket Deshpande, Lucky N Kapoor, Shivangi Kamat
*Department of Electrical and Electronics Engineering, Birla Institute of Technology and Science, Pilani, India 333031*

Satyanarayana Bheesette
*Department of High Energy Physics, Tata Institute of Fundamental Research, Mumbai, India 400005*

Dipankar Pal
*Department of Electrical and Electronics Engineering, Birla Institute of Technology and Science, Pilani, India 333031*

## Abstract

The paper presents the design and qualification tests of an airborne experimental setup to determine cosmic ray-flux in the lower stratospheric regions of earth's atmosphere. The concept of "coincidence" is implemented to preferentially detect cosmic rays and reject noise and particles that are incident at large angles but otherwise have similar characteristics, and are therefore inseparable from the particles of interest by conventional detection techniques. The experiment is designed to measure cosmic ray flux at two altitudes extending to a maximum height of 30 km from mean-sea-level. The experimental setup is to be lifted using a High Altitude Balloon (HAB). The setup is designed and tested to withstand extreme temperature and pressure-conditions during the flight in the stratosphere. It includes a cosmic ray telescope, a data acquisition system, a power supply systems and peripheral sensors. In the present endeavor the payload design and results from qualification tests are included.

## I. INTRODUCTION

Cosmic rays consist mainly of beta particles, hydrogen nuclei, alpha particles and atomic nuclei of heavier elements. These are continuously incident upon earth and have their sources spread across the universe with the sun being one of them. The charged particles are deflected by the magnetic field of earth. Nevertheless, some of them are able to pierce through the magnetic field as well. When these particles collide with the molecules in the upper atmosphere of the earth, they produce lighter and short-lived particles such as pions, kaons etc. Some of these particles further decay subsequently to produce muons, neutrinos, photons and electrons[1]. These interactions lead to drastic loss in the intensities of these particles making their energies inconsequential on the surface of the earth.

Radiation from cosmic rays is known to cause various anomalies in the electronic devices[1]-[6]. Ionizing particles of cosmic ray showers are known to mutate and damage cells in human tissues to even cause cancer[7]. Astronauts, aviation crews, frequent fliers are continually exposed to the high energy cosmic rays at high altitudes.

While astronauts are provided with protective equipment, detailed study and analyses is required to understand the implications of cosmic ray exposure to commercial and military aircraft crew and passengers [8]. The experiment, titled 'Project Apeiro' [9], aims to conduct measurement and analyses of the cosmic ray flux in lower stratospheric region of earth's atmosphere.

The experiment aims to obtain the cosmic ray dose rates by deploying an experimental setup as a payload on a High Altitude Balloon (HAB) in the lower stratospheric region. The setup is designed to measure cosmic ray dose rates by implementing the technique of 'coincidence' to identify cosmic ray particles. A single cosmic ray detector is highly sensitive, prone to noise and faulty triggers due to in-situ radiation, electronic noise among others. Therefore, the 'coincidence' technique presents a simple yet effective solution for distinguishing cosmic ray particles from noise.

Within the dimensions of the experimental setup, the cosmic ray particles are assumed to travel in straight lines, ignoring any curvature in its path. The general zenith angle distribution of incident cosmic rays on the earth can be denoted by:

$$I = I_o \cos^2 \theta \qquad (1)$$

where $\theta$ denotes the zenith angle of the cosmic ray particle path. Hence, cosmic rays can be presumed to be nearly perpendicular when they are incident upon earth's surface. The ambient charged particles on the other hand move in arbitrary directions and do not follow the vertical path. This characteristic of vertical incidence of cosmic rays is exploited in the proposed work to advantage for their selective detection in presence of other signal-generating sources. A set of particle detectors stacked vertically, one directly above another, senses a near vertical incident cosmic ray by producing signals simultaneously. Uniform output signals from all the detectors are therefore indicative of a cosmic ray particle while signals not coincident in all the detectors are taken as background noise and discarded.

This paper presents the design of an air-borne, complete cosmic-ray-detection-system payload and the results of the flight qualification tests of the payload. The rest of the paper is structured as follows. Section II presents the cosmic ray telescope description. Section III describes the data acquisition system for the telescope. In Section IV, low voltage and high voltage power supplies of the systems are detailed while Section V illustrates the protective measures and qualification as well calibration tests, undertaken for the high altitude balloon (HAB)-flight. Section VI concludes the paper.

## II. COSMIC RAY TELESCOPE

### A. Detector

A cosmic ray particle detector is constructed by connecting a Photomultiplier Tube (PMT) - Hamamatsu R6233 - to a plastic scintillator block of 100mm × 100mm using optical glue that maintains a consistent refractive index across the coupling. The PMT is operated at a high DC-voltage of 1000V, which is generated by an onboard DC-DCHV converter. A PMT based drive circuit, shown in FIG. 1 is used to bias the PMT dynodes. The entire assembly is covered with a thick sheet of black paper (called 'Tedler') to prevent stray light from outside entering the sensitive detector. A photograph of a fully assembled detector is shown in Fig 2. This detector assembly is then inserted inside an aluminum enclosure to protect it from electromagnetic interference.

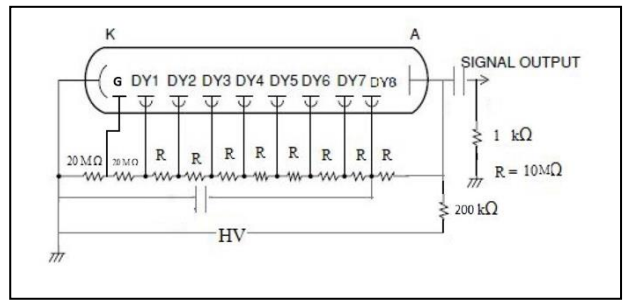

FIG. 1. Base circuit for the photomultiplier tubes

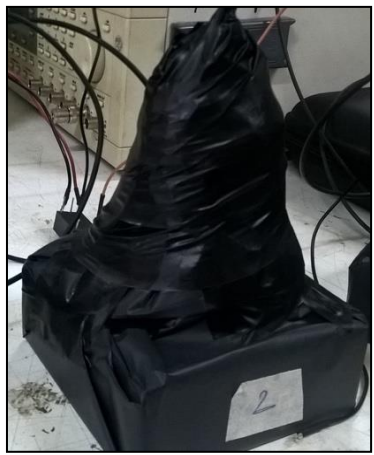

FIG. 2. A fully assembled single scintillator detector

### B. Telescope Arrangement

Three scintillation detectors are constructed and as stated in the previous section, stacked vertically one above the other, in order to obtain 'coincidence' of signals by all three scintillator detectors. This geometrical arrangement provides an aperture of 46.81° and can detect over 49% of the incident cosmic ray particles, distributed as in equation (1). The mechanical diagram of the arrangement is shown in FIG. 3. FIG. 4 is an actual photograph of the cosmic ray telescope. FIG. 5 is a schematic that explains how a cosmic ray shower passes through all the detectors while an in-situ charged particle passes through a single detector. Analog signals from each of the three scintillation detectors are sent to the data acquisition system for processing.

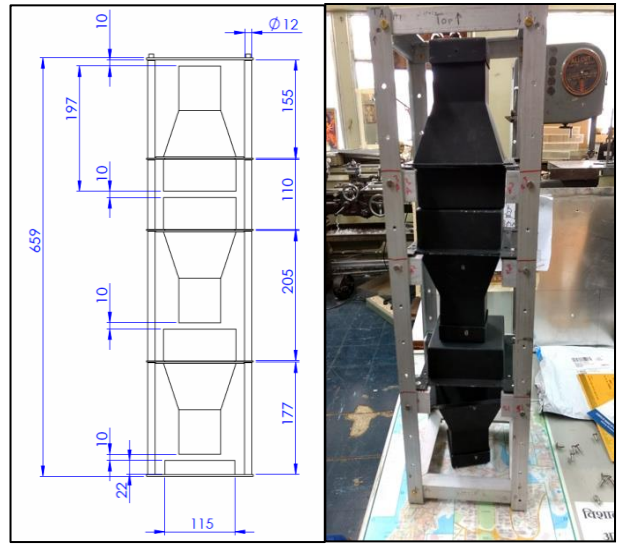

FIG. 3. (left) Mechanical drawing of the cosmic ray telescope and
FIG. 4. (right) Picture of the cosmic ray telescope

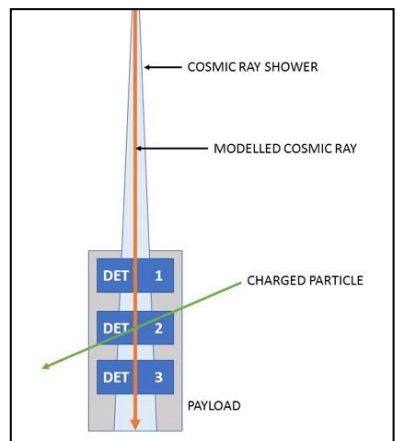

FIG. 5. Schematic of arrangement of scintillation detectors

## III. DATA ACQUISITION SYSTEM

*A. Overview*

The data acquisition system coverts negative polarity analog signals into logic signals using leading edge discriminators and also counts their individual rates. It performs logical coincidence of signals from all three detectors and then counts them. As stated, the coincidence-signals indicate passage of comic ray muons through the apparatus. The entire data acquisition system, which includes discrimination, coincidence and counting operations is implemented on a single printed circuit board (PCB). The DAQ system operations are controlled by a micro-controller. FIG.6 shows a functional blocks of the data acquisition system. Actual photograph of the same is shown in FIG. 7.

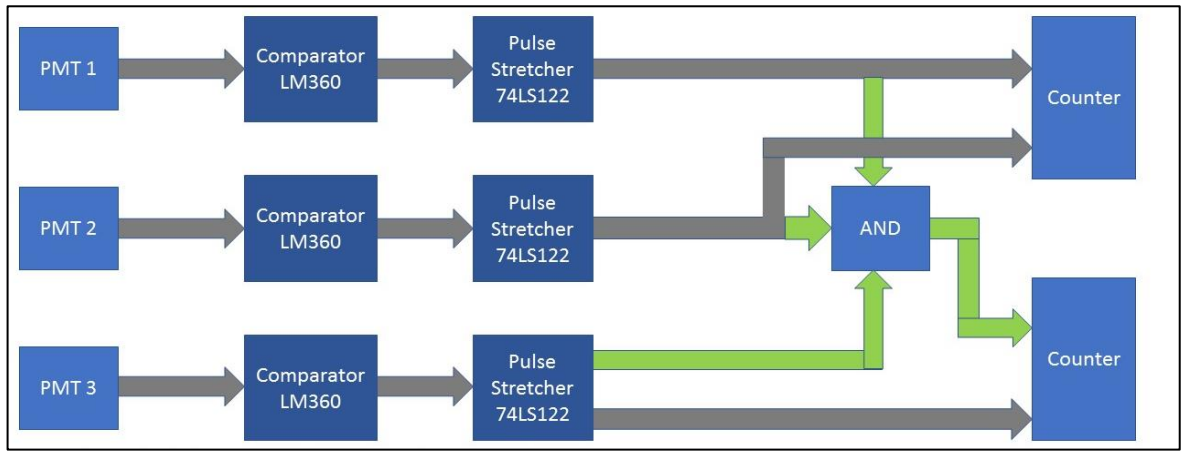

FIG. 6. Functional block diagram of the data acquisition system

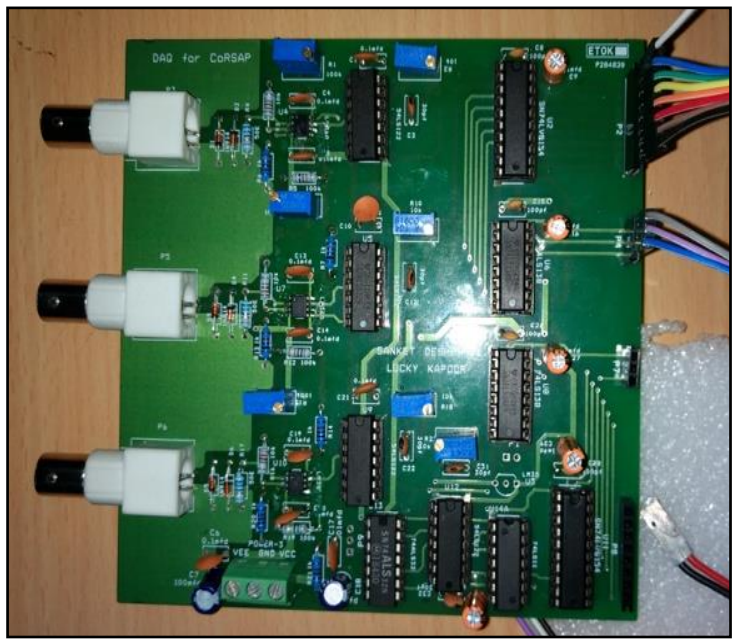

FIG. 7. Photograph of the data acquisition system board

*B. Discrimination*

The first step towards handling the incoming signals is to convert them into logic signals. The analog signals from the PMTs are in the form of fast pulses of negative polarity. These pulses are passed to a comparator with selectable threshold- LM360, which is a high frequency differential comparator that is chosen for this application. It generates a TTL-pulse of duration equal to the time that the analog pulse's magnitude is larger than the threshold. This way the comparator also discriminates or filters out signals of lower-than-the-threshold amplitude, which are essentially produced by the background radiation. Three discriminator channels are used.

SN54LS122J is a multivibrator, which is chosen to shape each TTL-pulse to a pulse of fixed duration for all channels. The time-duration is so set that it is greater than the maximum recorded time duration of the detector in a calibration experiment, yet smaller than the average time difference between two consecutive pulses from the same detector. The time duration of multivibrator pulses are adjusted by choosing a suitable resistor-capacitor combination. These signals are then individually sent to the counter ICs which count the number of signals of each detector for a selected time interval.

*C. Coincidence*

The coincidence circuit determines if there are simultaneous signals seen in all three detectors. A select logic is also provided to select the detectors which should participate in the coincidence logic. This provides flexibility to choose between three-fold and two-fold coincidences between and among the available detectors. The select logic is implemented using 2-input OR gates. Each OR gate receives one input each from the multivibrator and from the select line. The outputs of the OR gates are fed into a 3-input AND gate which performs the actual coincidence operation. If a select line is set to LOW, then the signal from the corresponding multivibrator is selected for coincidence. If the select line is set to HIGH, then the corresponding multivibrator signal is excluded from the coincidence operation. FIG. 8 is a block diagram which depicts selectable coincidence operation.

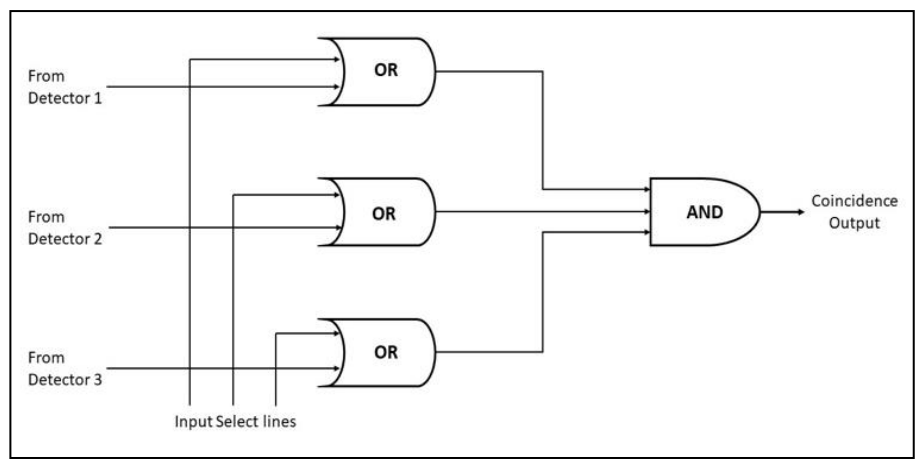

FIG. 8. Block diagram of the selectable detector coincidence circuit

*D. Counting*

The number of pulses from each of the multivibrators is detected and counted over a time period of 1 second. Therefore, four different counts are measured, namely, pulses form detector 1, pulses from detector 2, pulses from detector 3 and the coincidence pulses.

A combination of SN74LV8154 counter IC coupled with SN74LVC138A demultiplexer IC is used. The SN74LV8154 is a dual counter IC with inbuilt register, hence two such assemblies of counter-demultiplexer are required to perform counting operation on four signals. The IC has two 16-bit counters, whose output can be read over eight output lines. A select logic is implemented to obtain the upper eight bits or lower eight bits on the output lines. The demultiplexer has three chip select lines and three address lines. The input signals to the counter ICs are labeled as CLKA and CLKB.

The two sets of counter-demultiplexers are controlled and read simultaneously by the microcontroller. Therefore, there are sixteen data lines (eight for each IC. The chip select and address lines of both the demultiplexers are common. A typical readout operation follows the following sequence:
- Save counter values to inbuilt register
- Reset the counters
- Read lower bit of CLKA of both ICs
- Read upper bit of CLKA of both ICs
- Read lower bit of CLKB of both ICs
- Read upper bit of CLKB of both ICs

*E. Micro-controller, sensors and data handling*

The Texas Instruments MSP430F5438 Experimenter Board is designated as the controller for the payload. Besides providing counting rates of individual scintillator detectors as well as coincidences, the payload is designed to also record internal temperature and pressure, polar and azimuthal angular inclinations, GPS information, store data in an onboard EEPROM, receive commands from the ground station and continuously transmit data over the telemetry system to the ground station.

The temperature and pressure are measured using a BMP180 digital sensor module. The polar and azimuthal angles are measured using a MPU-9250 module while the GPS measurements are obtained via a Garmin 18x LVC module. Spansion S25FL216K 16MB EEPROM is used to store the experimental data on board. Commands from the ground station include inputs to the select lines for the coincidence logic and switching command for the high voltage power supply system. These are digital commands with TTL logic levels obtained from the tele-command unit over a buffered channel.

On powering up, the microcontroller executes the instructions to initialize the interfaced sensors and control ports, and also verifies every sensor by reading the pre-assigned, hexadecimal identification number. The microcontroller then reads the calibration registers of the BMP180 and MPU-9250 modules to note the initial configuration of the payload and its environment. After completing the initialization and calibration process, the microcontroller starts to acquire data.

The data acquisition, storage and transmission are executed in this sequence inside an always 'True' while-loop. The counts are read from the counter ICs by following the sequence mentioned in section D. The data from all peripherals are then collected. All the data acquired in a single run is compiled and stored in the on-board memory and also transmitted over the telemetry channel to the ground station. The counter ICs are read every second and the data is stored in a register. The telemetry data is sent at intervals of 2-seconds. The counts from two consecutive readouts are summed and transmitted over the telemetry channel, and also stored in the on-board EEPROM memory.

The telemetry data consists of experimental and housekeeping data. The experimental data is collected from all relevant systems and sensors, and is transmitted over a serial 3.3V level channel to the encoder. The packet-length is kept to a maximum of 120 characters with a baud rate of 9600 and refresh rate of 0.5 Hz. The housekeeping data is essential to monitor the health of the payload. This data includes voltages monitored at crucial points such as input voltage of High Voltage Power Supply (HVPS), 1000:1 scaled output voltage of HVPS, and positive and negative input voltages to the DAQ system. The housekeeping data is provided as analog buffered voltages to the telemetry encoding system. A block diagram of the sensors and their interfaces to the payload subsystems is given in FIG. 9.

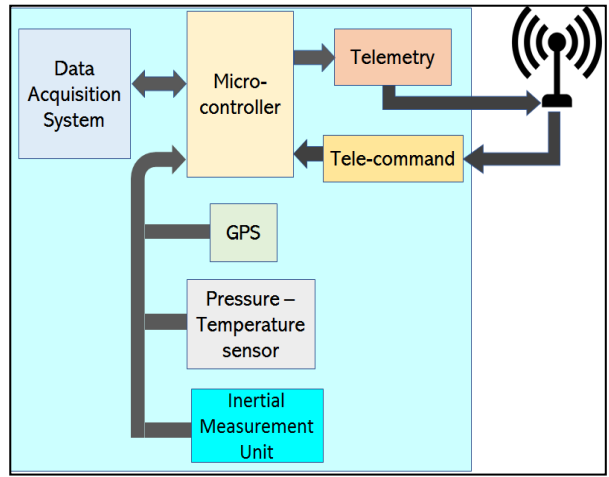

FIG. 9. Block diagram of sensors and interfacing

## IV. POWER SUPPLY SYSTEMS

The following types of power supplies are required by various subsystems of the payload:
1. 1000V DC: Photomultiplier Tubes (PMTs)
2. +5V DC: Majority of components operate at this voltage
3. -5V DC: Negative reference voltage for LM360
4. +3.3V DC: A few components operate at this logic level

Since the experimental setup is to be deployed on a HAB, the only power source available is Li-Po batteries. Power supplies are hence designed to utilize power from Li-Po batteries and to internally generate all the required voltages.

*A. High Voltage Power Supply (HVPS)*

The HVPS is designed to supply power to the PMTs. The HVPS circuit takes in 15V DC power as input which is fed to the linear power regulator IC LM7912 which produces an output of 12V. This voltage is provided to a voltage multiplier, 12AVR1000 by Pico Electronics, which produces an output voltage of 1000V DC. The ground-lines of the input and output to the voltage multiplier are kept isolated. The HVPS board's is shown in FIG. 10.

*B. Low Voltage Power Supply (LVPS)*

The LVPS board takes in 15V DC and -9V DC supplies as inputs and produces +5V DC, +3.3V DC and -5V DC outputs using linear power-supply-regulator ICs LM7805, LM1117 and LM7905 respectively.

## V. FLIGHT QUALIFICATION AND TESTING

The experimental setup is designed to be deployed on a HAB and sustain altitudes up to 30 km above mean sea level. The payload is expected to sustain low pressures of up to 20mbar and ambient temperatures lower than -50°C during its flight.

### A. Protective Measures

In low pressure environments, the exposed high voltage points are known to ionize the air molecules and hence are susceptible to electric arcing. Therefore, all the high voltage points were covered with a silicone elastomer Sylgard 170. The entire payload was covered with thermally-insulating polystyrene as a protection from ambient temperatures.

All circuits were individually enclosed inside aluminum casings to protect from the high EMI generated from the onboard HVPS circuits and DC-DC converters. All the interconnections between various subsystems were routed with a dual redundancy and via D-type connectors.

### B. PMT Characterization Test

The optimum operating voltage for the detector assembly was determined for accurate measurements. The setup included two qualified detectors with the test detector (to be denoted so in place of "Det 2" between them) as shown in Fig 3. The operating voltage of the test detector was varied from 600V to 1000V in steps of 25V. The two-fold coincidence between the two qualified detectors was compared to the three-fold coincidence to determine the efficiency. From FIG. 11, we can see that the efficiency for the detector saturates at 825V. Thus, voltage of 1000V was chosen as the operating voltage. This test was conducted for all the three PMTs and similar results were observed in all.

### C. Long Duration Power-On Test

In order to study the payload's integration with the flight instrument, the completely assembled equipment was subjected to a long duration test. In this test, the payload was kept in a room with controlled ambient temperature of about 24°C and was operated as expected in flight conditions. Data was acquired over telemetry and analyzed for possible errors. This test was conducted over a duration of over 16 hours.

Due to difference in construction of the detectors, the count rates varied slightly between them, but all were within acceptable ranges. As mentioned in Section III E, the counts are reported over intervals of two seconds over the telemetry channel. Over successive calibration tests, it has been observed that any count above 100, can be considered erroneous. Hence, all such readings were deleted. Moreover, there are a few instances wherein the telemetry sentence is distorted due to transmission issues. Such data points were also ignored from the analysis.

FIG. 12, shows a plot of detector counts against UTC time acquired during the long duration test. It can be observed that all the individual detector counts fell within the same range and the coincidence counts are generally lower than the individual detector counts. The system shows uniformity in operation over an extended duration of time and hence successfully cleared the test.

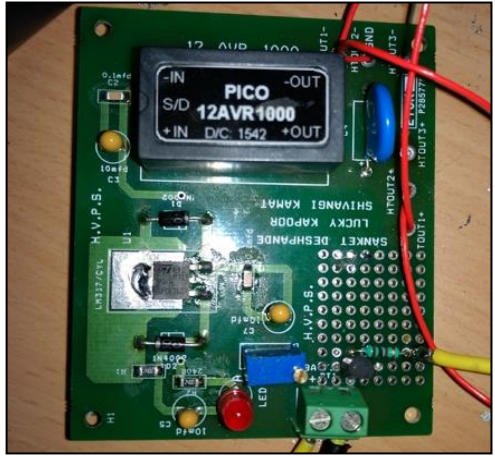

FIG. 10. Photograph of the High Voltage Power Supply unit

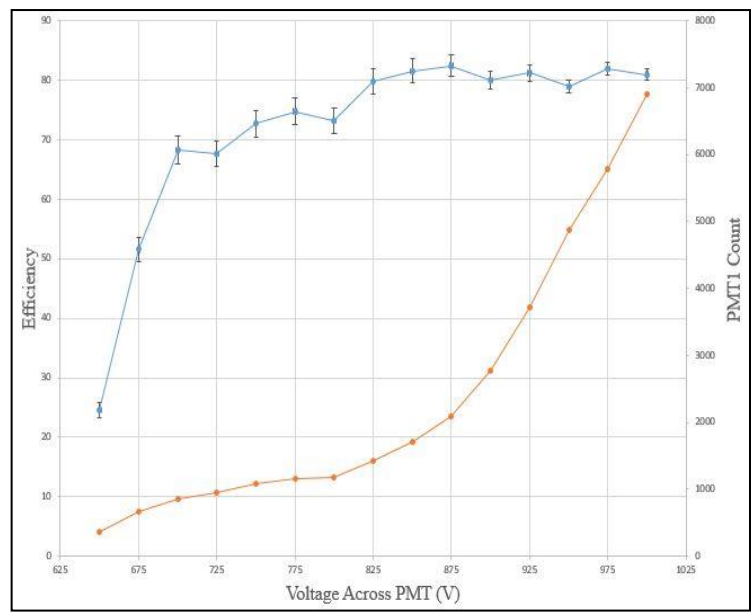

FIG.11. Efficiency vs Voltage plot for detector

*D. Flight Simulation Test*

The experimental setup was tested in an environmental chamber to simulate the actual flight conditions. The setup was not provided with any heat-insulation material.

The pressure was reduced to 20 mbar, corresponding to an approximate altitude of 30 km from mean sea level. The temperature was then reduced from 28ºC to -40ºC within a span of 70 minutes. The temperature and pressure were maintained at these values for 240 minutes for the payload to qualify for HAB flight.

All the parameters which were designed to be available during the flight via the telemetry channel were monitored during this test. This included the data over the serial channel, high voltage indicator and the coincidence logic selectors. Additionally, the following parameters were also monitored: input voltage to the HVPS board, input voltage to the LVPS board and input voltage to the DAQ board. Temperature probes were fitted on the payload for accurate measurement. A radiosonde was used to measure the ambient temperature and pressure for reference and cross-verification. The test conducted was successful.

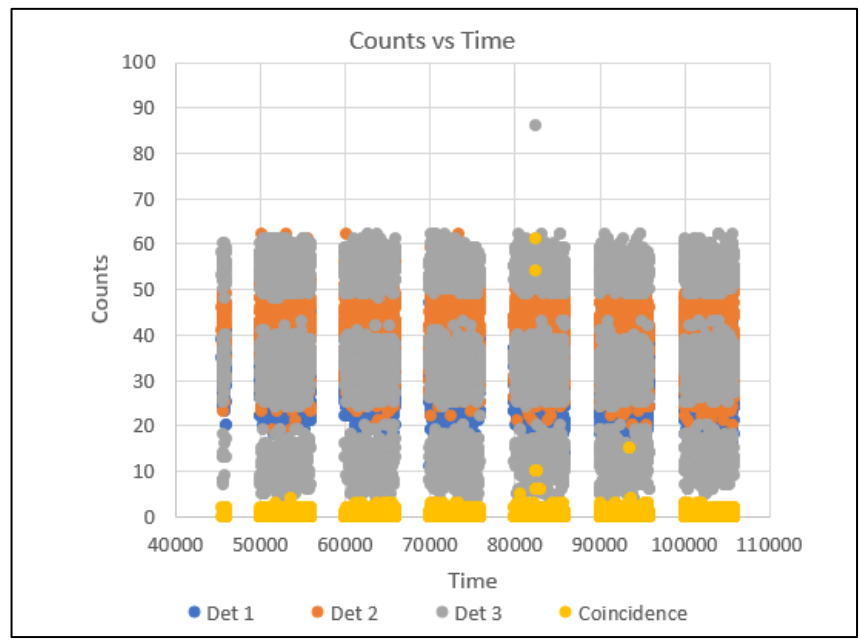

FIG. 12. Plot for detector counts vs UTC time

## VI. CONCLUSION

This paper presents the design and qualification tests followed for an experimental payload to determine cosmic ray flux rates in lower stratospheric regions. The system has been designed using commercially, off-the-shelf available ICs and components. The whole assembly was designed to be powered by lithium-ion batteries with circuit elements such as regulators and voltage multipliers used wherever high voltage was needed. Protective measurements and characterization test have been carried out and validated.

The system presented here is likely to outperform its expensive counterparts in design-simplicity without compromising on quality, novelty in idea of coincidence and discrimination apart.

## ACKNOWLEDGEMENTS

The authors are indebted to Tata Institute of Fundamental Research (TIFR) Dept. of High Energy Physics and TIFR Balloon Facility, Hyderabad for help in development and testing of the experimental setup. The authors would like to particularly acknowledge the efforts and support provided by Dr. Devendra Ojha (Chairperson, TIFR Balloon Facility), Mr. S. K. Buduru (Scientist-in-Charge, TIFR Balloon Facility), Mr. T. V. Rao (Flight Coordinator, TIFR Balloon Facility), Mr. B. V. N. Kapardhi, Mr. Santosh Koli, Mr. Dharmesh Trivedi, Dr. P. R. Sinha, Mr. Stalin Peter G, Mr. N. Nagendra, Mr. Manoj Kumar, Mr. R. R. Shinde, Mr. Yuvraj E, Mr. Pathaleshwar Esha and Mr. Vishal Asgolkar from TIFR.

The authors acknowledge the support of 'Project Apeiro' team members Pankaj Tiple, Vibhav Joshi and Srihari Menon and mentorship of Dr. Toby Joseph from BITS Pilani.